\documentclass[aoas,nameyear,seceqn,dvips]{arximspdf}
\usepackage{dcolumn}
\usepackage{graphicx}

\doi{10.1214/10-AOAS354}
\volume{4}
\issue{4}
\pubyear{2010}
\firstpage{2000}
\lastpage{2023}

\makeatletter

\newcommand{\eqref}[1]{(\ref{#1})}
\newcommand{\citeasnoun}[1]{\citet{#1}}

\newcommand{\bbeta}{\bolds\beta}
\newcommand{\btheta}{\bolds\theta}
\newcommand{\bOmega}{\bolds\Omega}

\newcommand{\bzero}{\textbf{0}}
\newcommand{\ba}{\textbf{a}}
\newcommand{\bD}{\textbf{D}}
\newcommand{\bI}{\textbf{I}}
\newcommand{\bu}{\textbf{u}}
\newcommand{\bw}{\textbf{w}}
\newcommand{\bX}{\textbf{X}}
\newcommand{\by}{\textbf{y}}
\newcolumntype{d}[1]{D{.}{.}{#1}}
\newcolumntype{a}[1]{D{,}{}{#1}}
\newcolumntype{b}[1]{D{.}{\,}{#1}}
\makeatother
\newtheorem{theorem}{Theorem}
\begin{document}
\begin{frontmatter}

\title{Generalized extreme value regression for binary response data:
An application to B2B electronic payments system adoption}
\runtitle{Generalized extreme value regression}

\begin{aug}
\author[A]{\fnms{Xia} \snm{Wang}\corref{}\ead[label=e1]{xiawang@niss.org}}
\and
\author[B]{\fnms{Dipak K.} \snm{Dey}\ead[label=e2]{dipak.dey@uconn.edu}}
\runauthor{X. Wang and D. K. Dey}
\affiliation{National Institute of Statistical Sciences and University
of Connecticut}
\address[A]{National Institute \\
\quad of Statistical Sciences\\
19 T.W. Alexander Drive\\
Research Triangle Park,\\
\quad North Carolina 27709\\USA\\ 
\printead{e1}}
\address[B]{Department of Statistics\\
University of Connecticut\\
215 Glenbrook Rd. U-4120\\
Storrs, Connecticut 06269\\USA\\
\printead{e2}}
\end{aug}

\received{\smonth{10} \syear{2009}}
\revised{\smonth{3} \syear{2010}}

%
\begin{abstract}
In the information system research, a question of particular interest
is to interpret and
to predict the probability of a firm to adopt a new technology such
that market promotions
are targeted to only those firms that were more likely to adopt the
technology. Typically, there exists significant difference between the
observed number of ``adopters'' and ``nonadopters,'' which is usually
coded as binary response. A critical issue involved in modeling such
binary response data is the appropriate choice of link functions in a
regression model. In this paper we introduce a new flexible skewed link
function for modeling binary response data based on the generalized
extreme value (GEV) distribution. We show how the proposed GEV links
provide more flexible and improved skewed link regression models than
the existing skewed links, especially when dealing with imbalance
between the observed number of 0's and 1's in a data. The flexibility
of the proposed model is illustrated through simulated data sets and a
billing data set of the electronic payments system adoption from a
Fortune 100 company in 2005.

\end{abstract}

%

\end{frontmatter}
%

\section{Introduction}\label{Introduction}

During the past three decades of information
system (IS) research, quite a rich but diverse body of theoretical and
empirical work has
accumulated on the adoption and diffusion of information technology
(IT) based innovations.
Technological advances and widespread uses of the internet are allowing
businesses to automate
a wide range of their business processes including payments. Electronic
payments system (EPS)
is an integrated process, in which payment data is sent and received
electronically from accounts
payable to accounts receivable without human intervention, and is a
critical component of the information economy.
Despite the tremendous benefits that EPS can offer, its adoption in
business-to-business (B2B) transactions remains
a challenge. In 2004, checks represented 81 percent of the typical
organizations' payments, according to the Association
for Financial Professionals (AFP) 2004 survey. A more recent survey by
AFP finds that, even though the payments are undergoing an
unprecedented period of change because of the decline of the check in
favor of electronic payments, a majority of B2B payments
continue to be made by check. The 2007 AFP survey indicates that the
typical organization is still making 74 percent of its
B2B payments by check. Researchers have tested different proposed
theories by showing the causal mechanism and identifying
influential factors. A number of barriers have been identified that
appear to obstruct the wider adoption of electronic payments
[\citeasnoun{Stavins2003}; \citeasnoun{ChakaravortiandDavis2004}].

Though exploring association based on the past data is critical to
understand the technology adoption behaviors, the ultimate objective of
EPS adoption studies would be accurate prediction as the field directly
aims at studying and solving practical problems. With no doubt,
accurate prediction is a key concern for practitioners since it is
anticipated future behavior that guides managerial action. The
immediate motivation of \citeasnoun{Bapnaetal2009}'s study, as quoted in
their paper, was a problem asked by the company, ``$\ldots$how can we
predict the likelihood of a firm to adopt EPS promoted by the company
so that we can target EPS promotions to only those firms that were
more likely to adopt EPS for their payments?'' Models with good
prediction accuracy are also needed by theorists in the field since
accurate prediction can serve as a direct ``reality check'' for the
relevance of the theoretical models, thus providing a base for theory
building, comparing and improving [\citeasnoun{ShmueliandKippius2009}]. A
good statistical model for EPS adoption study needs to provide not
only good fit to current data but also good prediction for future
observations.

Currently, the logistic regression model, with its convenient
interpretation and implementation, has been routinely employed to
estimate and predict the EPS adoption or other new technology adoption
in the literature [\citeasnoun{ChauandJim2002}; \citeasnoun{Bapnaetal2009};
\citeasnoun{GuptaandChintagunta1994}; \citeasnoun{KamakuraandRussell1989};
\citeasnoun{WedelandDeSarbo1993}]. When the logistic regression model is
employed, it is assumed that the response curve between the covariates
and the probability is symmetric. This assumption may not always be
true, and it may be severely violated when the number of observations
in the two response categories are significantly different from each
other. This unbalance is not uncommon in the IS field, since in many
cases the adoption of a new technology may be a ``rare event,'' which
happens with only a small probability. Thus, the data in this area is
usually complicated by the nature of the response variable being
analyzed: significant difference between the number of firms adopting
the technology and of those not adopting the technology and high
skewness in the response curve. Applying a nonflexible link function
to the data with this special feature may result in link
misspecification.

Consequences of link misspecification have been studied by a number of
authors in the literature. In particular, for independent binary
observations, \citeasnoun{CzadoandSantner1992} show that falsely assuming a
logistic link leads to a substantial increase in the bias and mean
squared error of the parameter estimates as well as the predicted
probabilities, both asymptotically and in finite samples. Moreover,
these undesirable effects are of greater magnitude when the
misspecification involves skewness than when it involves kurtosis (or
tail weight). \citeasnoun{WuChenandDey2002} show that under certain
conditions there exists linear relationships between the regression
coefficients, though the choice of links is important for goodness of
fit.

There has been considerable work done in allowing flexibility in the
link models used for categorical response data. The most intuitive
approach to guard against link misspecification is to embed the
symmetric links into a wide parametric class of links.
\citeasnoun{ArandaOrdaz1981} introduces two separate one-parameter models
for symmetric and asymmetric departures from the logistic model.
\citeasnoun{GuerreroandJohnson1982} suggest a one-parameter Box--Cox
transformation of the odds. \citeasnoun{Morgan1983} presents a
one-parameter cubic logistic model to model symmetric departures from
the logistic curve. It is a first-order approximation of the symmetric
Aranda-Ordaz model.

\citeasnoun{Stukel1988} proposes a class of generalized logistic models for
modeling binary data with a two-parameter family. Stukel's models are
general, and several important and commonly used symmetric and
asymmetric link models can be approximated by members of this family.
However, in the presence of covariates, Stukel's models yield improper
posterior distributions for many types of noninformative improper
priors, including the improper uniform prior for the regression
coefficients [\citeasnoun{ChenDeyandShao1999}]. Using a latent variable
approach of \citeasnoun{AlbertandChib1993}, \citeasnoun{ChenDeyandShao1999}
propose another class of skewed links, which can lead to proper
posterior distributions for the regression parameters using standard
improper priors. However, the model has the limitation that the
intercept term is confounded with the skewness parameter. This problem
was overcome in \citeasnoun{KimChenandDey2008} by a class of generalized
skewed $t$-link models (GT-link), though the constraint on the shape
parameter $\delta$ as $0<\delta\le1$ greatly reduces the possible
range of skewness provided by this model.

To build an appropriate and extremely flexible model for the binary
data and to overcome the constraint for the skewed generalized $t$-link
models, we propose the generalized extreme value (GEV) distribution as
a link function. We would first distinguish our GEV model from a line
of research on the discrete choice behavior initiated by
\citeasnoun{McFadden1978}. Although their model is also termed as
generalized extreme value models, it is totally different from the
model we introduce here. In their definition, GEV distribution is a
family of multivariate distribution functions whose marginal
distribution is Type I extreme value distribution or Gumbel
distribution [\citeasnoun{McFadden1978}], which is a special case of the
GEV distribution we use in $\eqref{GEV}$ with the shape parameter
$\xi\rightarrow 0$. Without a flexible shape parameter, their model
does not incorporate the skewness of the response curve but mainly
estimates the perceived similarity among alternatives
[\citeasnoun{McFadden1978}]. The advantage of the GEV link model we
discuss here is that it incorporates a wide range of skewness with the
unconstrained shape parameter. In fact, the complementary log--log
(Cloglog) link, based on the Gumbel distribution as discussed in
Section~\ref{Model}, is a special case of the proposed GEV link.

For model comparison, we use measures which have been suggested as a
criterion when the goal is to select a model with best out-of-sample
predictive power, including Deviance Information Criterion
(DIC) [\citeasnoun{Spiegelhalteretal2002}], Bayesian Information Criterion
(BIC) [\citeasnoun{KassandRaftery1995}] and marginal likelihood
[\citeasnoun{ChibandJeliazkov2001}; \citeasnoun{Chib1995}]. These three measures are
all in-sample predictive measures, which are computed using current
data. \citeasnoun{ShmueliandKippius2009} suggest that it is more general to
construct predictive accuracy measures based on out-of-sample metrics.
Thus, we also calculate posterior predictive deviance based on the
hold-out sample approach. Comparison of predictive performance also
guards against overfitting. Overfitting is a concern when building a
complicated and flexible model. Since an overfitting model
accommodates more of random fluctuations instead of the underlying
relationship, it will generally provide good fit to current data but
have poor predictive accuracy for new data. Good prediction results
provide evidence that overfitting has not occurred.

The rest of the paper is organized as follows. Section \ref{EDA}
describes the data set concerning the electronic system adoption of the
customers of a Fortune 100 company, where the dominance of
nonelectronic payment methods is particularly evident. Section
\ref{Model} describes the GEV model and discusses its posterior
propriety with Bayesian methodology. Section \ref{applications}
demonstrates the flexibility of the GEV model by simulated data sets
and the EPS data set described in Section \ref{EDA}. It concludes with
a discussion in Section \ref{Discussion}. The proof of the theorem
and detailed results on simulated data examples are given in the
\hyperref[appendix]{Appendix}. The computation for the GEV link described in this paper has
been implemented in R. A sample code [\citeasnoun{Wang2010}] is included
as supplementary material.

\section{The electronic payment system data set}\label{EDA}

To illustrate how the proposed GEV model may flexibly be used to model
and to predict the EPS adoption, we consider a billing data set from a
Fortune 100 companies (hereafter called the ``vendor''). It provides
information on individual transactions with firms who have financed
purchases of large commercial equipment with the vendor before 2005.
With a large amount to pay for the expensive commercial equipment,
firms usually finance the purchases with the vendor by lease or loan
ranging from a few months to more than 20 years. This brings in
subsequently recurring payments every month. The vendor provides all
client firms a voluntary free service of Automated Clearing House (ACH)
debits. Firms have the option to make their payments using ACH or
using conventional payment methods (such as a check). The data
contains payments choices of each transaction in the first quarter of
2005 as well as firm-specific and transaction-specific information,
including firm size, payment amount from the last bill, credit risk in
terms of PayDex, finance option, geographical regions and industries.
Firm size is measured by the total number of employees. The firm's
credit risk is based on a PayDex score, which is an indictor of a
business' payment performance evaluated on the total number of payment
experiences over the past year. Higher PayDex scores indicate better
payment performance. For each transaction, the firm has a financing
option between lease and loan. Region is a geographic variable, while
industry is a measure of firm demographics. Since these variables are
easy to observe and often associated with usage behavior, they have
been constantly used in marketing literature [\citeasnoun{Bapnaetal2009}].
EPS is the electronic payment methods adopted by the firm, with EPS $=0$
for ACH methods and EPS $=1$ for traditional methods such as check. We
have a total of 15,175 observations, with EPS $=1$ for 10,784
observations. That is, more than 70\% of the transactions used
traditional methods (EPS $=1$).
%
\begin{table}\tablewidth=240pt
\tabcolsep=0pt
\caption{Summary statistic for continuous variables} \label{EPScont}
\begin{tabular*}{240pt}{@{\extracolsep{\fill}}ld{2.1}d{5.2}d{7.1}@{}}
\hline
\textbf{Variable}& \multicolumn{1}{c}{\textbf{Min}} & \multicolumn{1}{c}{\textbf{Mean}} & \multicolumn{1}{c@{}}{\textbf{Max}} \\
\hline
\# of employees& 1 & 4142& 355000 \\
Payment amount& 11.4& 17351.0& 9520298.4 \\
PayDex & 5 & 71.83& 89 \\
\hline
\end{tabular*}
\end{table}
%
%
\begin{table}
\tablewidth=290pt
\tabcolsep=0pt
\caption{Summary statistic for categorical variables} \label{EPSdisc}
\begin{tabular*}{290pt}{@{\extracolsep{\fill}}la{5.0}a{5.0}d{2.4}@{}}
\hline
\textbf{Variables} &\multicolumn{1}{c}{\textbf{EPS $\bolds{=1}$}}& \multicolumn{1}{c}{\textbf{Counts}}& \multicolumn{1}{c@{}}{\textbf{\% EPS $\bolds{=1}$}} \\
\hline
Finance option      &      &       &       \\
\quad Lease         & 7857 & 10300,& 76.28\%\\
\quad\textit{Loan}       & 2927 & 4876, & \multicolumn{1}{a{2.1}@{}}{60\%}    \\
      Region        &      &       &          \\
\quad Midwest       & 3107 & 4608, & 67.43\%   \\
\quad Northeast     & 1466 & 2564, & 57.18\%    \\
\quad South         & 3688 & 4621, & 79.81\%     \\
\quad \textit{West} & 2523 & 3383, & 74.58\%      \\
Industry            &      &       &               \\
\quad Agriculture   & 127  & 256,  & 49.61\%        \\
\quad Construction  & 982  & 1556, & 63.11\%          \\
\quad Finance       & 308  & 416,  & 74.04\%         \\
\quad Manufacturing & 3340 & 4959, & 67.35\%        \\
\quad Mining        & 246  & 305   & 80.66\%       \\
\quad Public administration &318   & 327  &97.25\% \\
\quad Retail trade  & 396  & 69    & 84.44\%       \\
\quad Services      & 1599 & 2127  & 75.18\%      \\
\quad Transportation& 2320 & 3253  & 71.32\% \\
\quad \textit{Wholesale trade }& 1148 & 1508 & 76.13\%\\
EPS                 & 10784& 15175 &  71.06\%\\
\hline
\end{tabular*}
\end{table}

Summary statistics for this data set are reported in Tables
\ref{EPScont} and \ref{EPSdisc}. It provides a group of firms of
different sizes, from different industries and regions, and various
credit risk levels with different transaction amounts. As seen in
Table \ref{EPSdisc}, the proportions of EPS $=1$, which are almost all
higher than 50\%, vary within the subgroups based on finance option,
region or industry.

\begin{figure}

\includegraphics{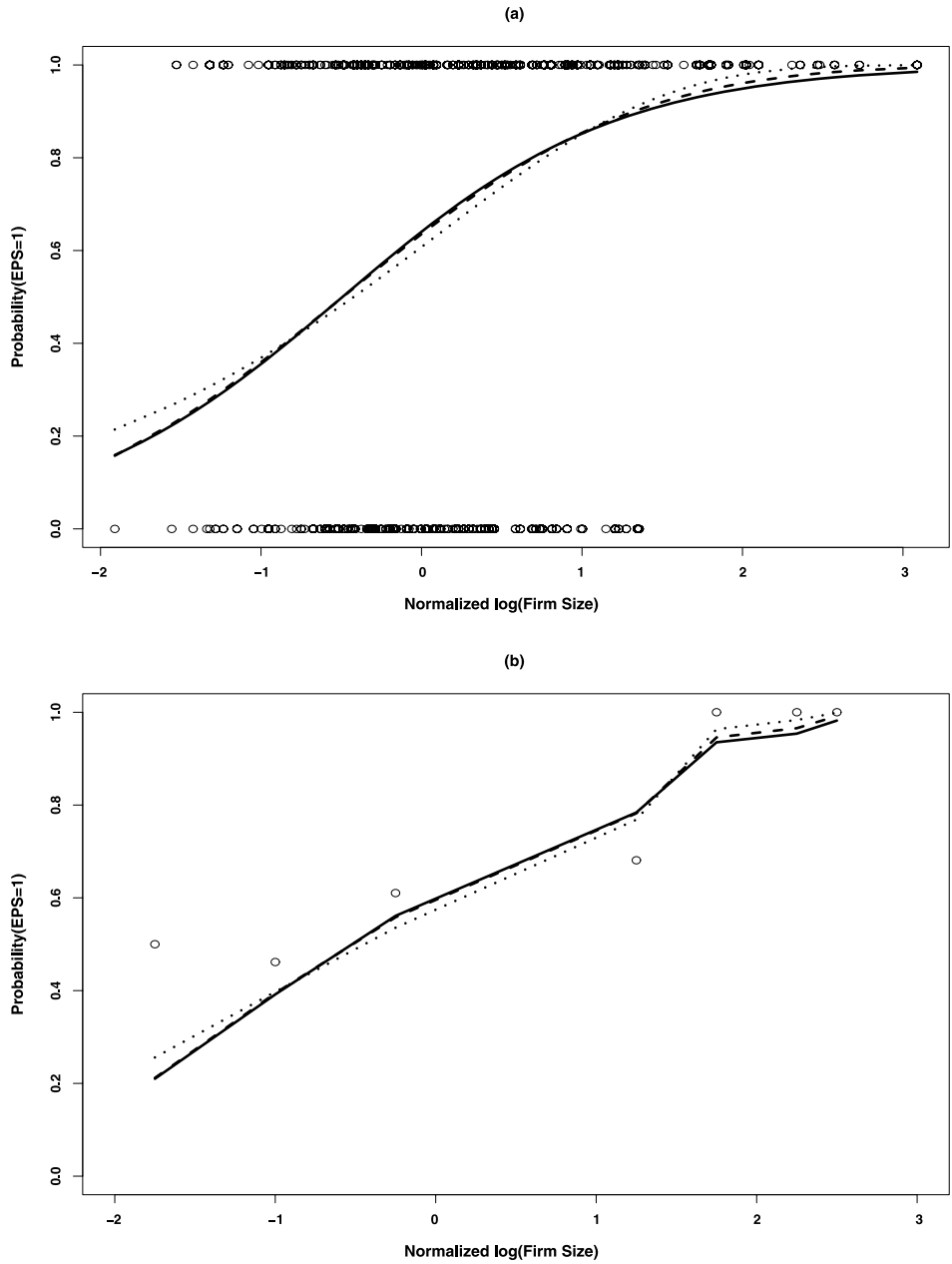}

\caption{Model fitting using the logit (solid), probit (dashed)
and Cloglog (dotted) link models to the subset of the EPS adoption data (circle) with
the normalized logarithm of firm sizes as a covariate: \textup{(a)} The original
data; \textup{(b)} collapse the firm sizes into 7 categories.}\label{MMWLdatafigure}
\end{figure}

We focus on a reduced sample to show the motivation of our proposed
model. The subset is selected from all transactions that involved
Midwest firms in the manufacturing industry with lease as the financing
option. This results in 1618 transactions. Out of these
transactions, 1222 transactions have EPS $=1$. We model the probability
of EPS $=1$ with the normalized logarithm of firm sizes as the only
convariate. This type of model would facilitate showing the response
curves under different link models in graphs. Figure
\ref{MMWLdatafigure}(a) shows the original data and the fitted
probability of the logit, probit and Cloglog model, respectively. To
view it more clearly, we collapse firm sizes into 7 categories as shown
in Table \ref{MMWLdatatable}. Also shown in Table \ref{MMWLdatatable}
are the estimated number of EPS $=1$ under different link models. Figure~\ref{MMWLdatafigure}(b)
 shows the observed values and the fitted values
under different link models. The positively skewed Cloglog link fits
the data better than the two symmetric links, though there is still
large discrepancy between the observed values and the estimated ones.
This is because the skewness inherited in the data is much higher than
that which can be provided by the Cloglog link, whose skewness is fixed
as a constant. A link model that can automatically change its shape of
the response curve based on the data would greatly improve the
estimation and prediction power of the model.

%

\section{Generalized extreme value link model}\label{Model} We first
specify the notation we use through the paper. Let $\by=(y_1, y_2,
\ldots, y_n)^{\prime}$ denote an $n \times1$ vector of $n$ independent
binary random variables. Also, let $\textbf{x}_i=(x_{i1}, x_{i2},
\ldots, x_{ik})^{\prime}$ be a $k \times1$ vector of covariates,
$i=1,\ldots,n$. Suppose $\bX$ denotes the $ n \times k$ design matrix
with rows $\textbf{x}_i^{\prime}$, and $\bbeta=(\beta_1, \beta_2,
\ldots, \beta_k)^{\prime}$ is a $k \times1$ vector of regression
coefficients. Assume that $y_i=1$ with probability $p_i$ and $y_i=0$
with probability $1-p_i$. In the generalized linear model framework,
%
\begin{equation}
p_i=\operatorname{Probability}(y_{i}=1)=F(\textbf{x}_i^{\prime}\bbeta), \label{link}
\end{equation}
where $F$ is a cumulative distribution function (cdf) and $F^{-1}$
determines the link function. The function
$F^{-1}(p_i)=\log\{p_i/(1-p_i)\}$ gives the logit link, which is a
symmetric link for binary response model. Another symmetric link model,
the probit link, is achieved by setting $F^{-1}(p_i)=\Phi^{-1}(p_i)$,
where $\Phi^{-1}$ is the inverse of $N(0,1)$ distribution. The
asymmetric Cloglog link is specified as
$F^{-1}(p_i)=-\log\{-\log(p_i)\}$.

%
\begin{table}
\tabcolsep=0pt
\caption{The estimated number of EPS $=1$ in the transactions with
Midwest manufacturing firms using lease as financing option under
different link models}\label{MMWLdatatable}
\begin{tabular*}{\textwidth}{@{\extracolsep{\fill}}lccccccc@{}}
\hline
\textbf{Normalized log(Firm size)} & \multicolumn{1}{c}{$\bolds{<-1.5}$}& \multicolumn{1}{c}{$\bolds{(-1.5, -0.5)}$}&\multicolumn{1}{c}{$\bolds{(-0.5,0)}$}
&\multicolumn{1}{c}{$\bolds{(0,1.5)}$}&\multicolumn{1}{c}{$\bolds{(1.5, 2)}$}&\multicolumn{1}{c}{$\bolds{(2, 2.5)}$}&\multicolumn{1}{c}{$\bolds{>2.5}$}\\
\hline
\# of EPS $=1$               &\multicolumn{1}{b{1.2}}{2.}    &\multicolumn{1}{b{2.2}}{90.}         &\multicolumn{1}{b{3.2}}{152.}
&\multicolumn{1}{b{3.2}}{410.}   &\multicolumn{1}{b{3.2}}{501.}    &\multicolumn{1}{b{2.2}}{47.}     &\multicolumn{1}{b{2.2}}{20.}\\
 Logit                    &0.84  &76.40        &139.73  &471.90 &468.56  &44.83   &19.64\\
 Probit                   &0.85  &76.90        &139.00  &470.89 &473.85  &45.39   &19.84\\
 Cloglog                  &1.02  &77.49        &133.36  &462.61 &482.75  &46.21   &19.98\\
\hline
\end{tabular*}
\end{table}
\subsection{The generalized extreme value distribution}
The $\operatorname{GEV}$ link models use the Generalized Extreme Value ($\operatorname{GEV}$)
distribution for $F$. Extreme value theory begins with a sequence of
independent and identically distributed random variables $Y_1, Y_2,
\ldots$ and, for a given $n$ asks about parametric models for its
maximum $M_n=\max\{Y_1, \ldots, Y_n\}$. If the distribution of the
$Y_i$ is specified, the exact distribution of $M_n$ is known. In the
absence of such specification, extreme value theory considers the
existence of $\lim_{n\rightarrow\infty}P[\{(M_n-b_n)/a_n\}\le
y]\equiv F(y)$ for two sequences of real numbers $a_n >0$ and
$b_n.$ If $F(y)$ is a nondegenerate distribution function, it belongs
to either the Gumbel, the Fr\'{e}chet or the Weibull class of
distributions, which can all be usefully expressed under the umbrella
of the $\operatorname{GEV}$ distribution with a cumulative distribution function as
follows:
%
\begin{equation}  \label{GEV}
G(x)=\exp\biggl[-\biggl\{1+\xi\frac{(x-\mu)}{\sigma}\biggr\}
^{-1/\xi}_{+}\biggr],
\end{equation}
where $\mu\in R$ is the location parameter, $\sigma\in R^{+}$ is the
scale parameter, $\xi\in R$ is the shape parameter and
$x_{+}=\max(x,0)$. A more detailed discussion on the extreme value
distributions can be found in \citeasnoun{Coles2001} and
\citeasnoun{Smith2003}. Extreme value analysis finds wide application in
many areas, including climatology [\citeasnoun{ColesPericchiandSisson2003};
\citeasnoun{SangandGelfand2009}], environmental science
[\citeasnoun{Smith1989}; \citeasnoun{Thompsonetal2001}], financial strategy of
risk management [\citeasnoun{DahanandMendelson2001};
\citeasnoun{ChamMorales2005}] and biomedical data processing
[\citeasnoun{Roberts2000}].

\begin{figure}

\includegraphics{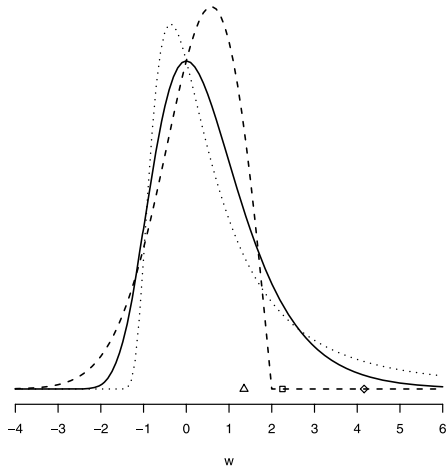}

\caption{Probability density function plots of Weibull
distribution ($\xi=-0.5$, dashed), Gumbel distribution ($\xi=0$, solid)
and Fr\'{e}chet distribution ($\xi=0.5$, dotted). The symbols
$\triangle, \Box$ and $\diamondsuit$ are the $90$th percentile of Weibull, Gumbel
and Fr\'{e}chet, respectively.}\label{GEVpdf}
\end{figure}

Its importance as a link function arises from the fact that the shape
parameter $\xi$ in model $\eqref{GEV}$ purely controls the tail
behavior of the distribution (see Figure~\ref{GEVpdf}). When
$\xi\rightarrow0$, it gives the Gumbel distribution with
$G(x)=\exp[-\exp\{-(x-\mu)/\sigma\}]$, which is the least positively
skewed distribution in the $\operatorname{GEV}$ class when $\xi$ is nonnegative.
%

\subsection{The generalized extreme value link model}
In model $\eqref{link}$ we assume that $F$ involves the
$\operatorname{GEV}$
distribution as follows:
%
\begin{equation} \label{(4)}
p_i=p(y_i=1)=1-\exp\{(1-\xi
\textbf{x}_i^{\prime}\bbeta)^{-1/\xi}_{+}\}
=1-\operatorname{GEV}(-\textbf{x}_i^{\prime}\bbeta;\xi),
\end{equation}
where $\operatorname{GEV}(x;\xi)$ represents the cumulative probability at $x$ for the
$\operatorname{GEV}$ distribution with $\mu=0$, $\sigma=1$, and an unknown shape
parameter $\xi$.

Since the usual definition of skewness in $\eqref{GEV}$ and
$\eqref{(4)}$ as
$\mu_3=\{E(x-\mu)^{3}\}\{E(x-\mu)^{2}\}^{-3/2}$ does not exist
for large positive values of $\xi$'s, we extend
\citeasnoun{ArnoldandGroeneveld1995}'s skewness measure in terms of the
mode to the $\operatorname{GEV}$ distribution for skewness definition and comparison.
Under certain conditions, the skewness of a random variable $X$ is
defined as $\gamma_M=1-2F(M_x)$, where $F(\cdot)$ is the cumulative
distribution of $X$ and $M_x$ is the mode of $X$. Thus, the skewness of
the link function $\eqref{(4)}$ can be found explicitly as
$\gamma_M=1-2F(M_x)=2\exp\{-(1+\xi)\}-1$, while $\xi>-1.$ Based on
this skewness definition and on the fact that for $\xi\leq-1$,
$\partial p/\partial x$ monotonically decreases with respect to $x$ in
$\eqref{(4)}$, we can show that the $\operatorname{GEV}$ link model specified in
$\eqref{(4)}$ is negatively skewed for $\xi<\log2-1$, and positively
skewed for $\xi>\log2-1 $.

\begin{figure}

\includegraphics{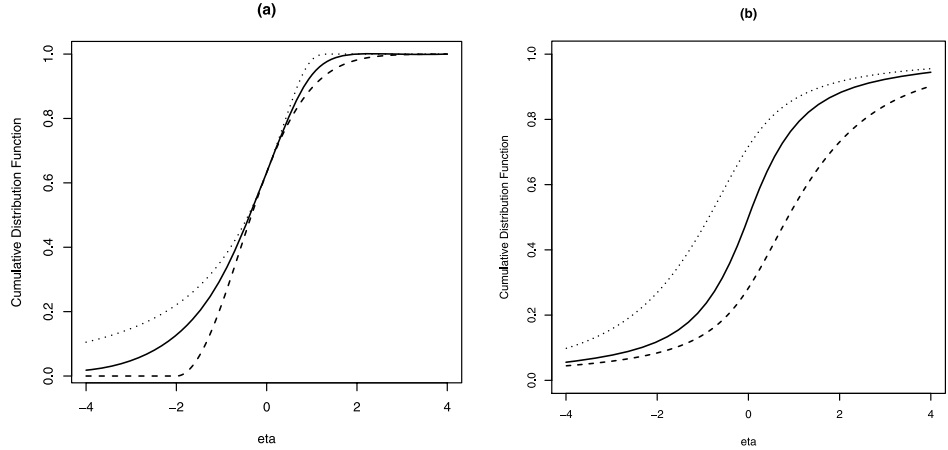}

\caption{Cumulative distribution function plots of \textup{(a)} the
$\operatorname{GEV}$
link with $\xi=-0.5$ (dashed), $0$~(solid) and $0.5$ (dotted), and \textup{(b)} the
GT link of $\delta z_i+\varepsilon_i$ with $v_1=1.2$, $\delta=1/v_2=1$,
and $G=\mathcal{E}$ (dotted) and $G=\mathcal{NE}$ (dashed), and the
solid curve is for $G=\triangle_{(0)}$.} \label{GEVGT}
\end{figure}

Figure \ref{GEVGT}(a) shows the response curves with $\xi$ equal to
$-0.5, 0$ and $0.5$. The solid line is the response curve
corresponding to the Cloglog link for $\xi\rightarrow0$. As the values
of the shape parameter change, so does the approaching rate to 1 and~0.
A much wider range of skewness can be fitted compared to the commonly
used Cloglog link. Figure \ref{GEVGT}(b) shows the distribution
function of the skewed generalized $t$-distribution of $\delta
z_i+\varepsilon_i$ with $v_1=1.2$, $\delta=1/v_2=1$
[\citeasnoun{KimChenandDey2008}]. The distribution $G$ in the GT link is
chosen as the standard exponential distribution ($\mathcal{E}$) and the
negative standard exponential distribution ($\mathcal{NE}$). The
skewness of the GT-link model is determined jointly by the constrained
$\delta$ and the preassumed $G$. Since $\delta=1$ is the maximum
possible value for the shape parameter, Figure \ref{GEVGT}(b) in fact
shows the two extreme cases of the GT-link with $G=\mathcal{E}$ or
$\mathcal{NE}$, whose skewness is 0.4339 for the $GT(\mathcal{E})$ case
and $-0.4339$ for the $GT(\mathcal{NE})$ case by
\citeasnoun{ArnoldandGroeneveld1995}'s measure. The range of skewness
provided by the $\operatorname{GEV}$ link models is not constrained. As shown in Figure
\ref{GEVGT}(a), even with $\xi\in[-0.5,0.5]$, the range of skewness
provided by the $\operatorname{GEV}$ links is much wider than that provided by the GT
links with a specified $G$.

%

\begin{figure}

\includegraphics{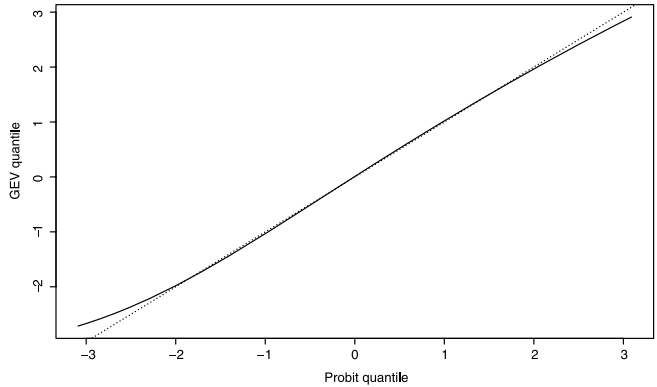}

\caption{Plot of $\operatorname{GEV}$ quantiles with $\mu\approx-0.35579,
\sigma\approx0.99903$, and $\xi\approx-0.27760$ against probit
quantiles for probabilities between 0.001 and 0.999. The solid line is
the quantile plot, and the dotted line is the $45^{\circ}$ reference
line.} \label{GEVprobit}
\end{figure}

The class of the $\operatorname{GEV}$ links also includes the symmetric link as a
special case. For example, by matching the first 3 moments, the
standard normal distribution can be approximated by the
$\operatorname{GEV}$
distribution with $\mu\approx-0.35579, \sigma\approx0.99903$, and
$\xi\approx-0.27760.$ Figure \ref{GEVprobit} shows the quantile
plots between the $\operatorname{GEV}$ model and the probit model. The plot is
approximately a straight line between 0.02 and 0.98 quantiles. The
discrepancy lies mainly in the tail area.
%

\subsection{Prior and posterior distributions for the generalized
extreme value link model}\label{Priors}
It is possible to estimate the shape parameter $\xi$ in the $\operatorname{GEV}$ link by
the maximum likelihood method. However, there are a number of
nonregular situations associated with $\xi$ in using likelihood methods
which requires conditions for usual asymptotic properties associated
with the maximum likelihood estimator to be valid. \citeasnoun{Smith1985}
studied the maximum likelihood estimation in nonregular cases in detail
and obtained that when $\xi<-0.5$ the regularity conditions are not
satisfied by $\operatorname{GEV}$ models. This violation of the usual regularity
conditions means that the standard asymptotic likelihood results are
not automatically applicable. This is one of the reasons for favoring a
Bayesian analysis since Bayesian methods do not depend on the
regularity assumptions required by the asymptotic theory of maximum
likelihood. In particular, in the unusual situation where $\xi<-0.5$
and the classical theory of maximum likelihood breaks down, Bayesian
inference provides a viable alternative. Thus, here we follow the
Bayesian methodology and fit the $\operatorname{GEV}$ link model in a general setting.

Let $\bD_{\mathrm{obs}}=(n, \by, \bX)$ denote the observed data. We assume that
the prior of $\xi$, $\pi(\xi)$, is proper. Then the joint posterior
distribution of $(\bbeta,\xi)$ based on $\bD_{\mathrm{obs}}$ is given by
%
\begin{equation}\label{post}
\pi(\bbeta,\xi|\bD_{\mathrm{obs}})\propto p(\by|\bX, \bbeta,
\xi)\pi(\bbeta|\xi)\pi(\xi),
\end{equation}
where $p(\by|\bX, \bbeta, \xi)
=\prod^n_{i=1}\{1-\operatorname{GEV}(-\textbf{x}_i^{\prime}\bbeta; \xi
)\}^{y_i}
\{\operatorname{GEV}(-\textbf{x}_i^{\prime}\bbeta; \xi)\}^{1-y_i}$
and\break
$\pi(\bbeta|\xi)$ is the conditional prior of $\bbeta$ given $\xi
$. In
Sections \ref{Simulation}--\ref{EPSwhole} we choose normal priors for
$\bbeta$ with $\beta_j \sim N (0,\sigma^2_{\beta_j})$,
$j=1,\ldots, k$, and for $\xi$ with $\xi\sim N (0, \
\sigma^2_{\xi})$, where priors on $\bbeta$ and $\xi$ are assumed
independent.

Unlike Stukel's generalized logistic regression model with covariates,
the posterior distributions under the $\operatorname{GEV}$ link are proper for many
noninformative priors, including the Jeffreys' prior and the improper
uniform prior for the regression coefficients $\bbeta$.

Jeffreys' prior for this model has the form $\pi(\bbeta|\xi)
\propto
|\bI(\bbeta|\xi)|^{1/2},$ where the Fisher information matrix
$\bI(\bbeta|\xi)$ is $\bX^{\prime}\bOmega\bX,$ with
$\bOmega=\mbox{diag} (\omega_1,\ldots, \omega_n),$ $\omega_i=\{
(1-\xi
\eta_i)^{-2/\xi-2}\}[\exp \{(1-\xi
\eta_i)^{-1/\xi}\}-1]^{-1}$, and $\eta_i =
\textbf{x}_i^{\prime}\bbeta,$ for $ i=1, \ldots, n.$ The joint
posterior is
then given by
\[
\pi(\bbeta,\xi|\bD_{\mathrm{obs}})\propto
\prod^n_{i=1}\{1-\operatorname{GEV}(-\textbf{x}_i^{\prime}\bbeta; \xi
)\}^{y_i}
\{\operatorname{GEV}(-\textbf{x}_i^{\prime}\bbeta; \xi)\}^{1-y_i}
|\bI(\bbeta|\xi)|^{1/2} \pi(\xi).
\]

The posterior propriety under the uniform prior established in Theorem
\ref{thm1}
 below implies that the proposed $\operatorname{GEV}$ link model is identifiable and
the estimation of the regression coefficients can contain little
subjective information in the Bayesian estimation. Let $\tau_i=-1$ if
$y_i=0$ and $\tau_i=1$ if $y_i=1$. Define $\bX^{*}_{l,m}=(\tau_i
\textbf{x}^{\prime}_i, l< i \leq m)$ as the $(m-l)\times k$ matrix with rows
$\tau_i \textbf{x}^{\prime}_i$, $l < i \leq m$, where $0 \leq l <m
\leq n$. We
are led to the following theorem concerning the propriety of the
posterior distribution in $\eqref{post}$ when $\pi(\bbeta) \propto1,$
which is an improper uniform prior and $\pi(\xi)=0.5$, for $
-1\leq\xi<1$, which corresponds to the uniform distribution $U[-1,1)$.
Even with this constraint, the $\operatorname{GEV}$ links are still fairly flexible with
skewness lying in the interval $[-0.7293, 1)$. The proof of Theorem \ref{thm1} is
given in the \hyperref[appendix]{Appendix}.

\begin{theorem}\label{thm1}
Suppose that there exist $p>k,$ $0=m_0 <
\cdots<m_p\leq n$, and positive vectors $\ba_1, \ldots, \ba_p$ such
that $\bX^{*}_{m_{l-1}, m_l}$ is of full rank and
$\ba^{\prime}_l\bX^{*}_{m_{l-1}, m_l}=\bzero$ for $l=1, \ldots, p$ and
with $\bzero$ a $k$-dimension zero vector. Under the improper uniform
prior $\pi(\bbeta) \propto1 $, the posterior $\eqref{post}$ is
proper.
\end{theorem}

\section{Applications of generalized extreme value link models}\label{applications}

\subsection{Model comparison criterion}\label{Comparison}
To assess models under different links, we employ four measures,
including Deviance Information Criterion (DIC), marginal likelihood,
Bayesian Information Criterion (BIC) and posterior predictive deviance
methods. The first three can be deemed as model adequacy measures
penalized by model complexity. Also, they measure the expected loss on
replicated data and can be used as model predictive power measures.
However, they are constructed based on current data
[\citeasnoun{ShmueliandKippius2009}]. Posterior deviance measures the
prediction accuracy directly based on hold-out samples. We briefly
discuss their computation and interpretation below.

Let $\btheta$ denote the set of all parameters contained in the model
under consideration. For example, $\btheta=(\bbeta, \xi)$ in the
$\operatorname{GEV}$
model. The deviance is defined as $-2$ times the log-likelihood, that
is, $D(\by,\btheta)=-2\log p(\by|\btheta).$ The posterior mean
deviance $\hat{D}_{\mathrm{avg}}$ serves as a Bayesian measure of fit or
``adequacy'' of a model [\citeasnoun{Spiegelhalteretal2002}], where
$\hat{D}_{\mathrm{avg}}(\by)=\{\sum^{L}_{l=1}D(\by,\btheta_{l})\}/L$ is the
estimated average discrepancy for $L$ sampling points and $\btheta_l$
is the $l$th sampling value for the parameter $\btheta$. The DIC
measure, which is proposed by \citeasnoun{Spiegelhalteretal2002}, is then
calculated as $\mathrm{DIC}=\hat{D}_{\mathrm{avg}}(\by)+p_{D}$, where $p_{D}$ is the
effective number of parameters of a Bayesian model. We calculated
$p_{D}$ as $\hat{p}=\hat{D}_{\mathrm{avg}}(\by)-D_{\hat{\btheta}}(\by)$, where
$D_{\hat{\btheta}}(\by)=D(\by,\hat{\btheta})$ and $\hat
{\btheta}$
is the posterior mean of the Markov chain Monte Carlo (MCMC)
simulation. The smaller the DIC value, the better the model fits the
data.

The marginal likelihood method is closely related to the Bayes factor,
which is given by $m(\by|M_{2})/m(\by|M_{1})$ with $m(\by|M_{i})$ the
marginal likelihood under model $M_i$, $i=1, 2$. The calculation of
the marginal likelihoods can be obtained by the MCMC method as
presented in \citeasnoun{Chib1995} and \citeasnoun{ChibandJeliazkov2001}.
\citeasnoun{ChibandJeliazkov2001}'s approach is more relevant here since
our models involve the Metropolis--Hastings algorithm.

The calculation of marginal likelihood is sensitive to the choice of
priors [\citeasnoun{KassandRaftery1995}], which may bring in potential
problems when the priors provide little information relative to the
information contained in the data. Under this situation, BIC is
usually applied as an approximation and it does not require evaluation
of the prior distributions [\citeasnoun{KassandRaftery1995}]. For a model
with $k$ parameters and $n$ observations, BIC is given by
$\mathrm{BIC}=-2\log\{p(\by|\hat{\btheta})\}+k\log(n)=D_{\hat{\btheta
}}(\by)+k\log(n).$

To assess the predictive power of the model, we aim at evaluating the
posterior predictive density $p(\tilde{\by}|\hat{\btheta}),$ where
$\tilde{\by}$ is the future response data (the holdout data set) and
$\hat{\btheta}$ are the posterior mean of the MCMC sampling from the
training data. The posterior predictive density can then be evaluated
by posterior deviance ($\hat{D}_{\mathrm{post}}$) as
$D(\tilde{\by},\hat{\btheta})=-2\log p(\tilde{\by}|\hat
{\btheta})$.

\subsection{Simulated data examples}\label{Simulation}
In the simulated data examples, we consider data sets simulated from
the Cloglog and probit regression models. Our primary aim is to show
the flexibility of the $\operatorname{GEV}$ link in fitting the data generated from
various models with unbalanced number of 0's and 1's. The true
parameters are set such that the proportion of 1's in the simulated
data sets is around 70\%, similar to the proportions in the AFP surveys
and the EPS data set. We perform Bayesian analysis for a given
simulated data set and assess the models using criteria stated in
Section \ref{Comparison}.

To match data scenario close to the EPS data, we generate 5 covariates
in our model, including the intercept. The types of covariates
represent those that occurred in the real data. It includes one
intercept ($\textbf{x}_1$), one continuous covariate generated from a standard
normal distribution ($\textbf{x}_2$) and three discrete covariates.
Among the
three discrete covariates, two are dummies for nominal categorical data
with 3 groups ($\textbf{x}_3$ and $\textbf{x}_4$) and the other is
binary categorical
data ($\textbf{x}_5$). All covariates are generated for sample sizes
$n=200,1000$ and $5000$, respectively. Then we generate two simulated
data sets with $n$ independent Bernoulli response variables, $y_i$,
respectively from (a) the Cloglog regression model as Simulation
$\sharp1$, and (b)~the probit regression model as Simulation $\sharp
2$. The linear components of all the above regression models are
$\textbf{x}_i^{\prime}\bbeta=\beta_1+x_{2i}\beta_2+x_{3i}\beta
_3+x_{4i}\beta_4+x_{5i}\beta_5$,
where $i=1, \ldots, n, \bbeta=(0, 1, 1, 0.5, -0.5)$ for the Cloglog
model and $(0, 1, 1, 1.25, -0.25)$ for the probit model. For all the link
models, we employ the Metropolis--Hasting algorithm with normal jumping
kernels for the MCMC sampling. The convergence of all results is
examined by the Bayesian Output Analysis (BOA) package in R.

Results are discussed in detail in the \hyperref[appendix]{Appendix} (Simulated Data
Examples). The $\operatorname{GEV}$ link performs better than the symmetric logit link
even with small sample size ($n=200$) under Simulation $\sharp1$. The
estimated $\xi$ includes the true value 0 in its $95\%$ high posterior
density (HPD) intervals. With increasing sample size, the variance of
$\xi$ decreases, which provides more precise estimation of the shape
parameter. In Simulation $\sharp2$, the $\operatorname{GEV}$ link approximates the
symmetric probit link well, especially with large sample size. The
impact of sample size on fitting the $\operatorname{GEV}$ link model is also reflected
by model selection based on BIC and marginal likelihood methods. These
two criteria tend to select a simpler model and the $\operatorname{GEV}$ link beats the
logit link only when $n=5000$ based on these two criteria under
Simulation $\sharp2$.

By the simulated examples, we also emphasize that the number of 1's and
0's is only an indicator of the possible skewness in the response
curve, though unbalance is not unusual even with a symmetric link
model, such as the probit link we used in Simulation $\sharp2$. As
suggested in \citeasnoun{ChenDeyandShao1999}, many factors, such as the
distribution of covariates, may affect the choice of links of a given
data set. This complexity is exactly why we propose a flexible link
function like the $\operatorname{GEV}$ link, such that the link function itself can
automatically detect and fit the symmetry or asymmetry in the response
curve along with parameter estimation. The probability of link
misspecification is reduced compared to applying a link model with a
fixed skewness parameter.

%
\begin{figure}

\includegraphics{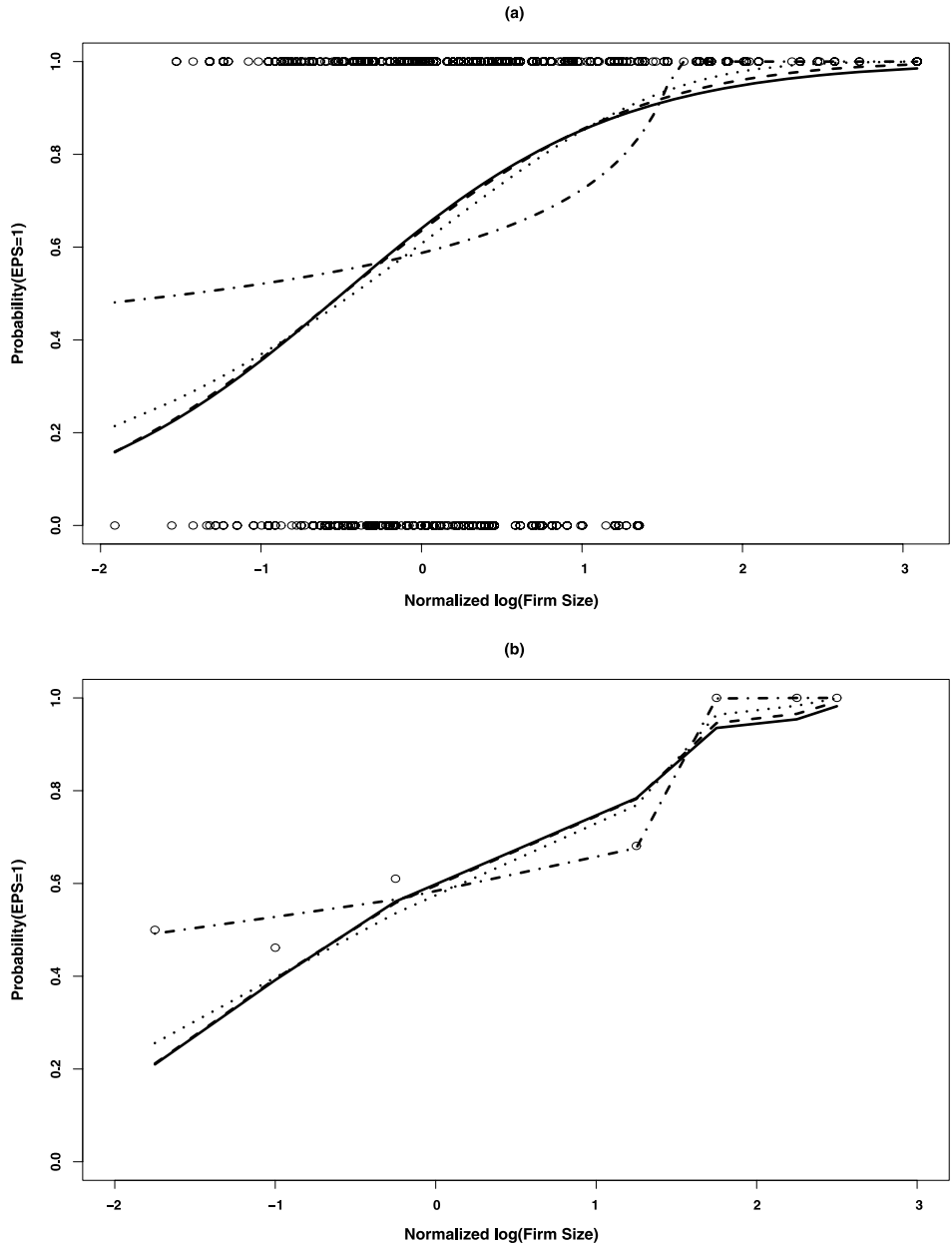}

\caption{Model fitting using the logit (solid),
probit (dashed), Cloglog (dotted) and GEV (dotdash) link functions to the
subset of the EPS
adoption data (circle)
with the normalized
logarithm of firm sizes
as a covariate. \textup{(a)} The original data; \textup{(b)} categorize firm sizes into 7
categories.} \label{MMWLdatafigure2}
\end{figure}

\subsection{The motivation subset of the electronic payments system
data}\label{EPSsubset}
We first fit the $\operatorname{GEV}$ model to the subset we discussed in Section
\ref{EDA}. Figure \ref{MMWLdatafigure2} shows (a) the fitted response
curve and (b) the fitted probability for the 7 categories based on the
firm's size. It is similar with Figure \ref{MMWLdatafigure} but with
the $\operatorname{GEV}$ link added. As shown in Figure \ref{MMWLdatafigure2}, the
response curve under the $\operatorname{GEV}$ link stretches significantly to fit the
observed values. Table \ref{MMWLdatatable2} shows the estimated number
of EPS $=1$, where we can see that the $\operatorname{GEV}$ model provides estimated values
that are very close to the observed ones. Also, we carry out the cross
validation analysis with randomly selected 10\% of the data as the
holdout part and the remaining data as the training part. The holdout
part has 173 transactions with 135 transactions EPS $=1$. The number of
observed EPS $=1$ in each of the 7 categories based on firm sizes is shown
in Table~\ref{MMWLdatatable3}. Here we can predict the number of EPS $=1$
in each category using the posterior mean estimates obtained from the
training part. It enables us to tell the predictive power of different
models more directly than the posterior deviance measure. The $\operatorname{GEV}$ link
still outperforms the other models and provides very good prediction
for those firms with normalized logarithm sizes greater than~0.

%
\begin{table}
\tabcolsep=0pt
\caption{The estimated number of EPS $=1$ in the transactions with
Midwest manufacturing firms using lease as financing option under
different link models} \label{MMWLdatatable2}
\begin{tabular*}{\textwidth}{@{\extracolsep{\fill}}lccccccc@{}}
\hline
\textbf{Standardized log(Firm size)} & $\bolds{<-1.5}$& $\bolds{(-1.5, -0.5)}$
&$\bolds{(-0.5,0)}$ & $\bolds{(0,1.5)}$ &$\bolds{(1.5,2)}$ & $\bolds{(2, 2.5)}$ &$\bolds{>2.5}$\\
\hline
\# of EPS $=1$ &\multicolumn{1}{b{1.2}}{2.} &\multicolumn{1}{b{3.2}}{90.} &\multicolumn{1}{b{3.2}}{152.} &\multicolumn{1}{b{3.2}}{410.}
&\multicolumn{1}{b{3.2}}{501.} &\multicolumn{1}{b{2.2}}{47.} &\multicolumn{1}{b{2.2}@{}}{20.}\\
 Logit &0.84 &\phantom{0}76.40 &139.73 & 471.90 &468.56 &44.83 &19.64\\
 Probit &0.85 &\phantom{0}76.90 &139.00 &470.89 &473.85 &45.39 &19.84\\
 Cloglog &1.02 &\phantom{0}77.49 &133.36 &462.61 &482.75 &46.21 &19.98\\
 $\operatorname{GEV}$ &\multicolumn{1}{b{1.2}}{2.} &102.96 &140.76 &407.10 &500.41 &47.00 &20.00\\
\hline
\end{tabular*}
\end{table}

%
%
\begin{table}
\tabcolsep=0pt
\caption{The predicted number of EPS $=1$ for transactions with
Midwest manufacturing firms using lease as financing option under
different links in the holdout part} \label{MMWLdatatable3}
\begin{tabular*}{\textwidth}{@{\extracolsep{\fill}}lccccccc@{}}\hline
\textbf{Standardized log(Firm size)} & $\bolds{<-1.5}$& $\bolds{(-1.5, -0.5)}$
&$\bolds{(-0.5,0)}$ & $\bolds{(0,1.5)}$ &$\bolds{(1.5,2)}$ & $\bolds{(2, 2.5)}$ & $\bolds{>2.5}$\\
\hline
\# of transactions &0 &\multicolumn{1}{b{1.2}}{18.} &\multicolumn{1}{b{2.2}}{22.} &\multicolumn{1}{b{2.2}}{64.} &\multicolumn{1}{b{2.2}}{62.}
&\multicolumn{1}{b{1.2}}{6.} &\multicolumn{1}{b{1.2}@{}}{1.} \\
\# of EPS $=1$ &--- &\multicolumn{1}{b{1.2}}{5.} &\multicolumn{1}{b{2.2}}{19.} &\multicolumn{1}{b{2.2}}{42.} &\multicolumn{1}{b{2.2}}{62.}
&\multicolumn{1}{b{1.2}}{6.} &\multicolumn{1}{b{1.2}@{}}{1.}\\
 Logit &--- &7.20 &12.20 & 49.71 &57.93 &5.74 &0.97\\
 Probit &--- &7.23 &12.13 &49.57 &58.58 &5.81 &0.98\\
 Cloglog &--- &7.23 &11.63 &48.64 &59.73 &5.91 &1.00\\
 $\operatorname{GEV}$ &--- &9.37 &12.25 &42.92 &61.94 &6.00 &1.00\\
\hline
\end{tabular*}
\end{table}

%
\subsection{The electronic payment system data set}\label{EPSwhole}
To further illustrate the flexibility of the proposed $\operatorname{GEV}$ link models,
we apply the model to analyze the whole EPS data. The aims of the
analysis are to examine the effects of various factors on EPS adoption
and to evaluate the fitness of regression models under different link
functions. The continuous variables, firm size and the payment amount,
are very skewed to the right. We take the logarithm and standardize
them by the sample mean and standard deviation. The credit risk by
Paydex is simply standardized. Financing option is a binary variable
with 1 indicating that the firm uses lease for financing its purchase
and 0 indicating loan. We have three dummy variables for the four
regions, and 9 dummy variables for the 10 industries, taking the values
0 and 1. Thus, there are a total of 16 covariates in the data for the
response EPS.

Table \ref{EPSest} shows the parameter estimation, its standard
deviation (SD), the average covariate effects (ACE), marginal
likelihood, BIC-16000, DIC-16000 and $\hat{p}$ for different link
functions using the normal priors. The prior variances for $\bbeta$'s and
$\xi$ are equal to $10^4$. A factor of 10 changes in these variance
settings led to almost identical posterior results.

We obtain $\hat{\xi}=1.40$ with a standard deviation of 0.079 for the
$\operatorname{GEV}$ link model, which indicates that $\xi$ is significantly above 0. In
fact, the value of DIC-16000 of the $\operatorname{GEV}$ link model is 446.68 with the
effective dimension $\hat{p}=17.30$, which is lower than 660.79 of the
Cloglog link model with $\hat{p}=16.91$. Both skewed link models are
better than the symmetric logit and probit link models. The consistent
model comparison results are obtained by using the marginal likelihood
and BIC criteria.

%
\begin{table}
\tabcolsep=0pt
\caption{Model comparison under logit, probit, Cloglog and
GEV links} \label{EPSest}
\fontsize{8}{10}\selectfont{
\begin{tabular*}{\textwidth}{@{\extracolsep{\fill}}ld{2.2}d{1.2}d{2.3}cd{2.2}d{1.2}d{2.3}cd{2.2}d{1.2}d{2.3}cd{2.2}d{2.2}d{2.3}@{}}
\hline
&\multicolumn{3} {c}{\textbf{Logit}} &&\multicolumn{3}{c}{\textbf{Probit}}&&\multicolumn{3}{c}{\textbf{Cloglog}}&&\multicolumn{3}{c}{$\bolds{\operatorname{GEV}}$}\\ [-7pt]
&\multicolumn{3}{c}{\hrulefill}&&\multicolumn{3}{c}{\hrulefill}&&\multicolumn{3}{c}{\hrulefill}&&\multicolumn{3}{c@{}}{\hrulefill}\\
\multicolumn{1}{@{}l}{\textbf{Variables}}
&\multicolumn{1}{c}{\textbf{est.}}&\multicolumn{1}{c}{\textbf{SD}}&\multicolumn{1}{c}{\textbf{ACE}}
&&\multicolumn{1}{c}{\textbf{est.}}&\multicolumn{1}{c}{\textbf{SD}}&\multicolumn{1}{c}{\textbf{ACE}}
&&\multicolumn{1}{c}{\textbf{est.}}&\multicolumn{1}{c}{\textbf{SD}}&\multicolumn{1}{c}{\textbf{ACE}}
&&\multicolumn{1}{c}{\textbf{est.}}&\multicolumn{1}{c}{\textbf{SD}}&\multicolumn{1}{c@{}}{\textbf{ACE}}\\
\hline
Intercept &0.97&0.08&&&0.60&0.05&&&0.24&0.04&&&0.13&0.03&\\
Firm size
&0.34&0.02&0.021&&0.21&0.01&0.020&&0.23&0.01&0.023&&0.17&0.01&0.024\\
Payment amount
&-0.17&0.02&-0.003&&-0.11&0.01&-0.003&&-0.11&0.01&-0.003&&-0.06&0.01&-0.002\\
Credit risk
&0.12&0.02&-0.029&&0.07&0.01&-0.028&&0.08&0.01&-0.034&&0.07&0.01&-0.039\\
Financing option &&&&&&&&&&&\\
\quad Lease
&0.44&0.04&0.086&&0.27&0.03&0.086&&0.27&0.03&0.088&&0.22&0.02&0.095\\
Region &&&&&&&&&&&\\
\quad Mid west
&-0.57&0.06&-0.109&&-0.34&0.03&-0.109&&-0.32&0.03&-0.103&&-0.13&0.02&-0.059\\
\quad North east
&-0.85&0.06&-0.177&&-0.53&0.03&-0.173&&-0.54&0.04&-0.181&&-0.42&0.03&-0.162\\
\quad South
&0.15&0.06&0.024&&0.08&0.03&0.028&&0.06&0.03&0.019&&-0.01&0.01&-0.005\\
Industry type &&&&&&&&&&&\\
 \quad Agri., forest, fish
&-1.15&0.14&-0.243&&-0.70&0.09&-0.242&&-0.80&0.10&-0.261&&-0.99&0.18&-0.272\\
 \quad Consturction
&-0.16&0.08&-0.029&&-0.09&0.05&-0.030&&-0.10&0.05&-0.033&&-0.17&0.03&-0.071\\
 \quad Fin., ins., reale.
&-0.07&0.13&-0.011&&-0.03&0.08&-0.014&&-0.04&0.07&-0.012&&-0.05&0.04&-0.023\\
 \quad Manufacturing
&-0.25&0.07&-0.044&&-0.14&0.04&-0.047&&-0.13&0.04&-0.041&&-0.08&0.02&-0.034\\
 \quad Mining
&0.42&0.16&0.057&&0.20&0.09&0.070&&0.10&0.08&0.031&&-0.06&0.04&-0.024\\
 \quad Public adm.
&2.58&0.36&0.255&&1.38&0.16&0.257&&1.10&0.12&0.248&&0.48&0.05&0.194\\
 \quad Retail trade
&0.43&0.15&0.069&&0.24&0.08&0.073&&0.18&0.07&0.054&&0.02&0.03&0.007\\
 \quad Service
&0.25&0.08&0.045&&0.15&0.05&0.045&&0.13&0.04&0.041&&0.04&0.02&0.017\\
\quad Trans., ele., gas&0.14&0.07&0.027&&0.09&0.05&0.025&&0.09&0.04&0.028&&0.01&0.02&0.003\\ [5pt]
Shape parameter ($\xi$) &&&&&&&&&&&&&1.40&0.08&\\ [5pt]
Marginal likelihood&\multicolumn{3}{d{5.2}}{-8489.2}   &&\multicolumn{3}{d{5.2}}{-8485.2}&&\multicolumn{3}{d{5.2}}{-8450.14}           &&\multicolumn{3}{d{5.2}}{-8359.24}\\
BIC-16000&\multicolumn{3}{d{5.2}}{888.71} &&\multicolumn{3}{d{5.2}}{862.63}&&\multicolumn{3}{d{5.2}}{790.64}&&\multicolumn{3}{d{5.2}}{585.38}\\
DIC-16000 &\multicolumn{3}{d{5.2}}{759.45}&&\multicolumn{3}{d{5.2}}{732.84}&&\multicolumn{3}{d{5.2}}{660.79}&&\multicolumn{3}{d{5.2}}{446.68}\\ [5pt]
$\hat{p}$ &\multicolumn{3}{d{5.2}}{17.21}&&\multicolumn{3}{d{5.2}}{16.94}&&\multicolumn{3}{d{5.2}}{16.91}&&\multicolumn{3}{d{5.2}}{17.30}\\
\hline
\end{tabular*}
}
\end{table}

A closer look at the results from the widely used logit link regression
model in the IS research and our $\operatorname{GEV}$ regression model reveals some
difference in the estimation of the covariates' effects on EPS
adoption. The changes in covariates include doubling the firm size,
doubling the payment amount, 10 points decrease in the paydex measure,
or moving from 0 to 1 for all the discrete covariates. The average
covariate effects are calculated based on
\citeasnoun{ChibandJeliazkov2006}. A~point to clarify is that the changes
in the firm size, payment amount and the credit risk are on their
original levels, while the regression itself has been run with these
covariates standardized. The logit model suggests that transactions by
firms in the southern region may have a lower probability employing
EPS compared to those by firms in the western region with 1.9--2.8\%
changes in the probability, but the estimation from the $\operatorname{GEV}$ model
implies that there is no significant difference in these two regions
with covariate effect approximately equal to a $0.54\%$ increase in
the probability. Also, the $\operatorname{GEV}$ model indicates that transactions by
firms in the construction industry are significantly more likely to use
EPS ($7\%$ higher in probability) compared with those by firms in the
wholesale trade industry, while the logit model suggests that it is not
a statistically significant factor with the estimated average covariate
effects around $3\%$. Even when a parameter is significant in both
models, the impact of changes in the covariate may be quite different.
For example, firms in the Midwest are approximately $11\%$ more likely
to use check compared to those in the west based on the logit model,
while the probability is only around $6\%$ as estimated by the
$\operatorname{GEV}$
model.

To calculate the posterior deviance of the four different link models,
we randomly divide the data into training and hold-out parts, with 10\%
or 1501 observations as the hold-out part and we calculate the
posterior deviance ($\hat{D}_{\mathrm{post}}$), which are 1653.40, 1643.04,
1637.92 and 1618.73 for logit, probit, Cloglog and $\operatorname{GEV}$ links,
respectively. The $\operatorname{GEV}$ link model, with the lowest $\hat{D}_{\mathrm{post}}$,
outperforms the other three commonly used models in prediction. Thus,
there is no indication of overfitting in the proposed model.

Since the assignment of 0 and 1 to the response variable is arbitrary,
we can reverse the role of 0 and 1. It does not affect the fitness of
the symmetric links, but it changes the asymmetry in the data. If we
now define EPS $=1$ if the transaction used the ACH service and 0 if it
used the conventional payment method, we expect that a positively
skewed link model like the Cloglog link is not appropriate anymore. In
fact, we fitted the data using the same MCMC sampling method, and the
values of DIC-16000 for the logit, the probit, the Cloglog and the
$\operatorname{GEV}$
link are 759.03, 733.1, 815.49 and 516.66, respectively. The Cloglog
fits the data poorly, just as expected, since it is a positively skewed
link function and the response function is now negatively skewed. This
result shows that the determination of links should not just be a
matter of choosing between a symmetric link and a skewed link, and that
the direction of a skewed link indeed plays a more important role,
since a wrong choice of direction gives an even worse fit than a wrong
choice between symmetric and asymmetric. The flexibility of the
$\operatorname{GEV}$
link models is apparent here as it includes the shape parameter as a
model parameter to be estimated through the data, instead of
preassuming any direction of it. With the redefined 0 and 1 in the
response variable, the $\operatorname{GEV}$ link model has a negatively significant
shape parameter of $-1.11$ with a standard deviation of 0.05.

\section{Conclusion and discussion}\label{Discussion} This paper
introduced a new flexible skewed link model for analyzing binary
response data with covariates in the EPS adoption setup. The proposed
$\operatorname{GEV}$ link model provided great flexibility in fitting skewness in the
response curve. Although theoretically the GT link models in \citeasnoun
{KimChenandDey2008} would be rich enough to have similar flexibility,
it is not easily implementable from a computational perspective.
Moreover, the computation burden is much less for the $\operatorname{GEV}$ link. For a
simulated data with 2 covariates, it took approximately 2 minutes to
fit the $\operatorname{GEV}$ model in R, while it took approximately 8~hours to fit the
GT model with codes in FORTRAN 95.

One existing problem in the $\operatorname{GEV}$ link is that the shape parameter $\xi$
also affects the heaviness of the tail in the distribution. Its
flexibility would be further improved if we can design a mechanism to
modify the $\operatorname{GEV}$ distribution such that one parameter would purely serve
as skewness parameter while the other could purely control the
heaviness of the tails.
\begin{appendix}\label{appendix}
\section*{Appendix}
\subsection*{Proof of Theorem \protect\ref{thm1}}
Let $u, u_1, \ldots, u_n$ be independent random variables with common
distribution function $F$, which is a $\operatorname{GEV}$ distribution with $\mu=0$,
$\sigma=1$, and a shape parameter $\xi$. For $0<a<1$, it can be shown
that $E|u|^{a}<\infty$ for $-1\leq\xi< 1$. Observing that $1-F(-x)=E
I(u>-x)$ and $F(-x)=E I\{-u \leq-(-x)\},$ here $I$ is an indicator
function. Now, we have
$\{1-F(-\textbf{x}_i^{\prime}\bbeta)\}^{y_i}\{F(-\textbf
{x}_i^{\prime}\bbeta)\}^{1-y_i}\leq
E I\{\tau_i u_i \geq\tau_i (-\textbf{x}_i^{\prime} \bbeta)\}$ and
$\{1-F(-\textbf{x}_i^{\prime}\bbeta)\}^{y_i}\times\break\{F(-\textbf
{x}_i^{\prime}\bbeta)\}^{1-y_i}\geq
E I\{\tau_i u_i > \tau_i (-\textbf{x}_i^{\prime}\bbeta)\} .$ Let
$\bu^{*}=(\tau_1u_1, \ldots, \tau_n u_n).$ Using Fubini's theorem, we
obtain
{\makeatletter
\def\@eqnnum{\hb@xt@.01\p@{}\rlap{\normalfont\normalcolor
                 \hskip -\displayindent\hskip -\displaywidth\theequation}}
\def\theequation{(A.1)}
\makeatother
\begin{eqnarray}\label{A1}
&&\int^{1}_{-1}\int_{R_k}p(\by|\bX, \bbeta, \xi)\,d\bbeta\,d\xi\nonumber\\
&&\qquad = \int^{1}_{-1}\int_{R_n}E\biggl[ \int_{R_k}I\{-\tau_i \textbf{x}_i^{\prime}\bbeta<\tau_i u_i, 1\leq i\leq n\}\,d\bbeta\biggr]\,dF(\bu)\,d\xi\\
&&\qquad =
\int^{1}_{-1}\int_{R_n}E\biggl\{\int_{R_k}I(\bX^{*}\bbeta<\bu^{*})\,d\bbeta\biggr\}\,dF(\bu)\,d\xi.\nonumber
\end{eqnarray}
}

Under the condition of Theorem \ref{thm1}, it follows directly from Lemma 4.1 of
\citeasnoun{ChenandShao2000} that there exists a constant $K$ such that
$\Vert\bbeta\Vert \leq K \min_{1\leq l\leq p}\!(\max_{m_{l-1}<i\leq m_l}|w_i|)
$ whenever $\bX^{*}\bbeta\leq\bw,$ where $\bw=(w_1, \ldots, w_n).$
Hence, from \ref{A1}, we have
\begin{eqnarray*}
&&\int^{1}_{-1}\int_{R_k}p(\by|\bX, \bbeta, \xi)\,d\bbeta\,d\xi\\
&&\qquad\leq K\int^{1}_{-1}\int_{R_n}\prod_{l=1}^{k}E\Bigl({\max_{m_{l-1}<i\leq m_l}}|\bu^{*}|^{k/p}\Bigr)\,dF(\bu)\,d\xi<\infty.
\end{eqnarray*}

\subsection*{Simulated data examples}
We fit Cloglog, $\operatorname{GEV}$ and logit models for Simulation $\sharp1$. The
$\operatorname{GEV}$
model gives almost identical estimates of the regression coefficient
$\bbeta=(\beta_1, \beta_2, \beta_3, \beta_4, \beta_5)$ as the true
Cloglog regression model. The estimated shape parameter and its $95\%$
high posterior density intervals are $0.402 (-0.199, 1.193), 0.058
(-0.269, 0.175)$ and $-0.034 (-0.136, 0.073)$ as $n=200, 1000$ and
$5000$. The true value of $\xi$ is already contained in the $95\%$ HPD
interval when $n=200$. However, the variance in the estimation of the
$\xi$ becomes smaller as the sample size increases, which implies that
the $\operatorname{GEV}$ links do require enough information contained in the data set
for more efficient estimation of the shape parameter.

\citeasnoun{CzadoandSantner1992} emphasized that it is more appropriate to
study the link misspecificaiton effects on the estimated probabilities
since estimation of $\bbeta$ heavily depends on the chosen link
function. They suggested that different link models can only be
\emph{unambiguously} compared in their estimation of event
probabilities. Table \ref{cloglog1} shows the average covariate
effects. These effects are calculated by the method suggested in
\citeasnoun{ChibandJeliazkov2006}, where they marginalize out the
covariates as a Monte Carlo average using their empirical distributions
and integrate out the parameter $\btheta$ by their posterior
distribution. The third column in Table \ref{cloglog1} indicates that
the average covariate effect is measured as $\textbf{x}_2$ doubled, or
$\textbf{x}_3, \textbf{x}_4$ or $\textbf{x}_5$ moving from 0 to 1.
The values from the
Cloglog model are in bold font. The average covariate effects
estimated by the $\operatorname{GEV}$ model are closer to those estimated by the Cloglog
model than the logit model.

%
\begin{table}\tablewidth=270pt
\tabcolsep=0pt
\caption{Simulation $\sharp 1$  based on the Cloglog regression model:
average covariate effect comparison} \label{cloglog1}
\begin{tabular*}{270pt}{@{\extracolsep{\fill}}lccrrr@{}}
\hline
&&\multicolumn{1}{c}{\textbf{Covariate}}&&&\\
\multicolumn{1}{c@{}}{\textbf{Sample size}} & \multicolumn{1}{c}{\textbf{Parameter}} &\multicolumn{1}{c}{\textbf{change}}
& \multicolumn{1}{c}{\textbf{Cloglog}}
&\multicolumn{1}{c}{$\bolds{\operatorname{GEV}}$}
&\multicolumn{1}{c}{\textbf{Logit}}\\
\hline
$n=200$ &$\beta_2$&double &\textbf{0.2071}\phantom{00}&0.2033\phantom{0}&0.1985\\
&$\beta_3$&$0 \rightarrow 1$&\textbf{0.2409}\phantom{00}&0.2533\phantom{0}&0.2191\\
&$\beta_4$&$0 \rightarrow 1$&\textbf{0.1100}\phantom{00}&0.1150\phantom{0}&0.0894\\
&$\beta_5$&$0 \rightarrow 1$&$\bolds-$\textbf{0.1332}\phantom{00}&$-$0.1428\phantom{0}&$-$0.1300\\[5pt]
$n=1000$ &$\beta_2$&double &\textbf{0.1902}\phantom{00}&0.1889\phantom{0}&0.1826\\
&$\beta_3$&$0 \rightarrow 1$&\textbf{0.1771}\phantom{00}&0.1737\phantom{0}&0.1624\\
&$\beta_4$&$0 \rightarrow 1$&\textbf{0.0861}\phantom{00}&0.0844\phantom{0}&0.0807\\
&$\beta_5$&$0 \rightarrow 1$&$\bolds-$\textbf{0.1306}\phantom{00}&$-$0.1330\phantom{0}&$-$0.1460\\[5pt]
$n=5000$ &$\beta_2$&double &\textbf{0.1925}\phantom{00}&0.1924\phantom{0}&0.1875\\
&$\beta_3$&$0 \rightarrow 1$&\textbf{0.2082}\phantom{00}&0.2096\phantom{0}&0.2152\\
&$\beta_4$&$0 \rightarrow 1$&\textbf{0.1016}\phantom{00}&0.1021\phantom{0}&0.1024\\
&$\beta_5$&$0 \rightarrow 1$&$\bolds-$\textbf{0.1122}\phantom{00}&$-$0.1115\phantom{0}&$-$0.1067\\
\hline
\end{tabular*}
\end{table}

%
\begin{table}\tablewidth=280pt
\tabcolsep=0pt
\caption{Simulation $\sharp1$ based on the Cloglog regression model:
model comparison} \label{cloglog2}
\begin{tabular*}{280pt}{@{\extracolsep{\fill}}lcrrr@{}}
\hline
& &\multicolumn{1}{c}{\hspace*{-3pt}\textbf{Cloglog}}&\multicolumn{1}{c}{\hspace*{-3pt}$\bolds{\operatorname{GEV}}$} &\multicolumn{1}{c@{}}{\textbf{Logit}} \\
\hline
$n=200$ &$D_{\hat{\btheta}}$&176.57\phantom{0}&175.64\phantom{0}&181.38\\
&DIC &186.24\phantom{0} &186.98\phantom{0}&191.47\\
&$\hat{p}$&4.84\phantom{0}&5.67\phantom{0}&5.05\\
&marginal likelihood&$-$119.34\phantom{0}&$-$120.96\phantom{0}&$-$118.31\\
&BIC&203.06\phantom{0}&207.43\phantom{0}&207.87\\
&$\hat{D}_{\mathrm{post}}$&168.65\phantom{0}&169.13\phantom{0}&173.34\\[5pt]
$n=1000$ &$D_{\hat{\btheta}}$&890.96\phantom{0}&890.69\phantom{0}&896.78\\
&DIC &901.06\phantom{0} &902.62\phantom{0}&906.99\\
&$\hat{p}$&5.05\phantom{0}&5.96\phantom{0}&5.10\\
&marginal likelihood&$-$481.02\phantom{0}&$-$486.37\phantom{0}&$-$481.17\\
&BIC&925.50\phantom{0}&932.14\phantom{0}&932.32\\
&$\hat{D}_{\mathrm{post}}$&863.69\phantom{0}&863.81\phantom{0}&872.52\\[5pt]
$n=5000$ &$D_{\hat{\btheta}}$&4388.53\phantom{0}&4388.19\phantom{0}&4427.16\\
&DIC &4398.52\phantom{0} &4399.88\phantom{0}&4436.85\\
&$\hat{p}$&5.00\phantom{0}&5.85\phantom{0}&4.84\\
&marginal likelihood&$-$2233.73\phantom{0}&$-$2240.97\phantom{0}&$-$2250.31\\
&BIC&4431.11\phantom{0}&4439.29\phantom{0}&4469.75\\
&$\hat{D}_{\mathrm{post}}$&4418.96\phantom{0}&4419.69\phantom{0}&4473.86\\
\hline
\end{tabular*}
\end{table}

Table \ref{cloglog2} shows the model comparison results. We also
include the effective number of parameters to show model complexity.
To test the predictive power of different link models, for each
simulated data set, we generate another data set of the same size $n\,(=200, 1000, 5000)$
 with the same $\bbeta$'s as the hold-out part and
the originally simulated data set as the training part. Based on the
deviance in Table \ref{cloglog2}, the $\operatorname{GEV}$ link outperforms both the
Cloglog model and the logit model. The DIC measure allows us to take
the model complexity into account. The DIC values for the $\operatorname{GEV}$ model are
very close to the Cloglog model and both the $\operatorname{GEV}$ model and the Cloglog
model provide better fit than the logit model based on the DIC measure.
This result is consistent with the fact that the Cloglog model is a
special case of the $\operatorname{GEV}$ model. Simulation $\sharp1$ is in fact based
on a $\operatorname{GEV}$ regression model with $\xi\rightarrow0$. The comparison is
consistent also in the aspect of the predictive power. In Table
\ref{cloglog2} the $\operatorname{GEV}$ model has almost the same posterior deviance as
the Cloglog model, while the logit model is worse than both models for
different sample sizes. Comparison based on the marginal likelihood and
BIC is only consistent with the above results when $n=5000$, which may
suggest that benefits applying a $\operatorname{GEV}$ model are more prominent with
large sample size. The better predictive performance of the $\operatorname{GEV}$ model
suggests that there is no overfitting by the proposed model.

%
\begin{table}\tablewidth=280pt
\tabcolsep=0pt
\caption{Simulation $\sharp2$  based on the probit regression model:
average covariate effect comparison} \label{probit}
\begin{tabular*}{280pt}{@{\extracolsep{\fill}}lccrrrr@{}}
\hline
&&\multicolumn{1}{c}{\textbf{Covariate}}&&& \\
\multicolumn{1}{c}{\textbf{Sample size}} & \multicolumn{1}{c}{\textbf{Parameter}}
&\multicolumn{1}{c}{\textbf{change}}
& \multicolumn{1}{c}{\textbf{Probit}} &\multicolumn{1}{c}{\textbf{Logit}}
&\multicolumn{1}{c}{\textbf{Cloglog}} &\multicolumn{1}{c}{$\bolds{\operatorname{GEV}}$}\\
\hline
$n=200$ &$\beta_2$&double &\textbf{0.1757}&0.1752\phantom{0}&0.1757\phantom{0}&0.1727\\
&$\beta_3$&$0 \rightarrow 1$&\textbf{0.1619}&0.1636\phantom{0}&0.15443&0.1579\\
&$\beta_4$&$0 \rightarrow 1$&\textbf{0.2867}&0.2861\phantom{0}&0.2871\phantom{0}&0.2821\\
&$\beta_5$&$0 \rightarrow 1$&$\bolds-$\textbf{0.0328}&$-$0.03478&$-$0.0123\phantom{0}&$-$0.0285\\
\\
$n=1000$ &$\beta_2$&double &\textbf{0.1892}&0.1882\phantom{0}&0.1925\phantom{0}&0.1906\\
&$\beta_3$&$0 \rightarrow 1$&\textbf{0.1756}&0.1766\phantom{0}&0.1838\phantom{0}&0.1778\\
&$\beta_4$&$0 \rightarrow 1$&\textbf{0.2630}&0.2641\phantom{0}&0.2512\phantom{0}&0.2583\\
&$\beta_5$&$0 \rightarrow 1$&$\bolds-$\textbf{0.0683}&$-$0.0650\phantom{0}&$-$0.0689\phantom{0}&$-$0.0703\\
\\
$n=5000$ &$\beta_2$&double &\textbf{0.1871}&0.1862\phantom{0}&0.1866\phantom{0}&0.1877\\
&$\beta_3$&$0 \rightarrow 1$&\textbf{0.2254}&0.2253\phantom{0}&0.2253\phantom{0}&0.2259\\
&$\beta_4$&$0 \rightarrow 1$&\textbf{0.2859}&0.2858\phantom{0}&0.2825\phantom{0}&0.2858\\
&$\beta_5$&$0 \rightarrow 1$&$\bolds-$\textbf{0.0420}&$-$0.0417\phantom{0}&$-$0.0403\phantom{0}&$-$0.0417\\
\hline
\end{tabular*}
\end{table}

The results from
Simulation $\sharp2$ are
summarized in Tables \ref{probit}
and \ref{probit2}. The main objective of Simulation $\sharp2$ is to show that the
$\operatorname{GEV}$
model can approximate a symmetric probit link model. The 95$\%$ HPD
intervals of the estimated shape parameter $\xi$ are $(-0.8,
0.17),(-0.36, -0.03)$ and $(-0.31, -0.13)$ for $n=200, 1000$ and
$5000$, which all include the value $\xi\approx-0.27760$ at which the
$\operatorname{GEV}$ distribution approximates the standard normal distribution. A more
precise estimation of $\xi$ does require larger sample size. The
sample size also has impact on how the $\operatorname{GEV}$ model compares with the
other models. As shown in Table \ref{probit2}, only at $n=5000$ all the
model comparison criteria except the marginal likelihood method agree
that the more complex $\operatorname{GEV}$ link model exhibits some advantages compared
to the logit and Cloglog links. The marginal likelihood method is
affected by the flat priors on parameters. It requires an even larger
sample size to select the $\operatorname{GEV}$ link as a better model.

%
\begin{table}\tablewidth=280pt
\tabcolsep=0pt
\caption{Simulation $\sharp2$ based on the probit regression model:
model comparison} \label{probit2}
\begin{tabular*}{280pt}{@{\extracolsep{\fill}}lcrrrr@{}}
\hline
& &\multicolumn{1}{c}{\textbf{Probit}}&\multicolumn{1}{c}{\hspace*{1pt}\textbf{Logit}}
&\multicolumn{1}{c}{\textbf{Cloglog}}
&\multicolumn{1}{c@{}}{\hspace*{2pt}$\bolds{\operatorname{GEV}}$} \\
\hline
$n=200$ &$D_{\hat{\btheta}}$&154.74&155.13&156.25&154.63\\
&DIC &164.56 &164.90 &166.10&165.96\\
&$\hat{p}$&4.91&4.89&4.92&5.67\\
&marginal likelihood&$-$108.23&$-$105.73&$-$109.25&$-$114.48\\
&BIC&181.23&181.62&182.74&186.42\\
&$\hat{D}_{\mathrm{post}}$&199.80&200.45&201.69&199.67\\
\\
$n=1000$ &$D_{\hat{\btheta}}$&861.44&862.70&865.28&861.02\\
&DIC &871.17 &872.64 &875.31&872.68\\
&$\hat{p}$&4.86&4.97&5.01&5.83\\
&marginal likelihood&$-$466.17&$-$463.97&$-$468.05&$-$472.90\\
&BIC&895.98&897.23&899.82&902.46\\
&$\hat{D}_{\mathrm{post}}$&843.24&844.42&851.27&844.60\\
\\
$n=5000$ &$D_{\hat{\btheta}}$&4219.46&4226.72&4238.51&4217.05\\
&DIC &4229.43 &4236.51 &4248.40&4228.71\\
&$\hat{p}$&4.99&4.89&4.94&5.83\\
&marginal likelihood&$-$2149.16&$-$2149.96&$-$2158.67&$-$2155.75\\
&BIC&4262.04&4269.31&4281.09&4268.15\\
&$\hat{D}_{\mathrm{post}}$&4190.66&4196.77&4228.71&4195.57\\
\hline
\end{tabular*}
\end{table}

\end{appendix}

\section*{Acknowledgments} The research of Xia Wang was part of her dissertation
at the Department of Statistics, University of Connecticut. The
research of Dipak K. Dey
was funded in part by SAMSI, Research Triangle Park, N.C., during his
visit as part of his sabbatical.
The authors wish to thank Dr. Zhongju Zhang for providing the data set.
The authors gratefully acknowledge
the insightful comments and constructive suggestions by the Associate
Editor and two anonymous referees.

\begin{supplement}[id=suppA]
\stitle{R codes for GEV models with covariates}
\slink[doi]{10.1214/10-AOAS354SUPP}
\slink[url]{http://lib.stat.cmu.edu/aoas/354/supplement.txt}
\sdatatype{.txt}
\sdescription{ The computation for the $\operatorname{GEV}$ link described in this
paper has been implemented in R which is available in this
supplementary material.}
\end{supplement}

\printaddresses
\end{document}